# Sleeper Social Bots: A New Generation of AI Disinformation Bots are Already a Political Threat

Jaiv Doshi[1], Ines Novacic[1], Curtis Fletcher[1], Mats Borges[1], Elea Zhong[1], Mark C. Marino[1], Jason Gan[1], Sophia Mager[1], Dane Sprague[1], and Melinda Xia[1]

[1]University of Southern California

## Abstract

This paper presents a study on the growing threat of "sleeper social bots," AI-driven social bots in the political landscape, created to spread disinformation and manipulate public opinion. We based the name sleeper social bots on their ability to pass as humans on social platforms, where they're embedded like political "sleeper" agents, making them harder to detect and more disruptive. To illustrate the threat these bots pose, our research team at the University of Southern California constructed a demonstration using a private Mastodon server, where ChatGPT-driven bots, programmed with distinct personalities and political viewpoints, engaged in discussions with human participants about a fictional electoral proposition. Our preliminary findings suggest these bots can convincingly pass as human users, actively participate in conversations, and effectively disseminate disinformation. Moreover, they can adapt their arguments based on the responses of human interlocutors, showcasing their dynamic and persuasive capabilities. College students participating in initial experiments failed to identify our bots, underscoring the urgent need for increased awareness and education about the dangers of AI-driven disinformation, and in particular, disinformation spread by bots. The implications of our research point to the significant challenges posed by social bots in the upcoming 2024 U.S. presidential election and beyond.

## 1. Introduction

Twenty twenty-four will see the first U.S. presidential election since generative AI technology became widely accessible. Bots, having evolved from spreading fake news in 2016 to promoting QAnon conspiracies in 2020, will continue to cause political chaos and confusion. With the addition of Large Language Models (LLMs) like ChatGPT, disinformation spread by bots will exponentially amplify political un-truths online, especially on social media where they can now more readily pass as humans. These new LLM-powered bots surpass the behaviors of their predecessors which simply posted easily identifiable automated messages. As a result, they can and will slip past user defenses and infiltrate networks on social media. For this reason, we call this new class **sleeper social bots.**

## 2. Introducing Sleeper Social Bots

In previous waves of bots, programs followed specific grammars, repeated patterns of phrase, or behavior that could be randomized or tied to other content, such as drawing upon headlines from a news source. These bots tended to be tied to a single topic about which they could post (Lampi, 2017). While any single post could mix into a stream of content and go largely undetected because it is difficult to filter out from others jumping on a re-posting bandwagon, watched closely, these bots easily revealed their automated underpinnings. What sets new bots apart is their ability to engage in unrehearsed, spontaneous dialogue with others. While previous bots were repetitive and primarily uni-directional, contemporary social bots leverage large language models (LLMs) to conduct conversations convincingly. This new class of bots draw from extensive training data to generate responses to new inputs, making their interactions seem more human (Li, Yang, and Zhao 2023). Trained on internet discussions, these bots are already versed in a vast array of political discourse.

The differences between these two classes of bots could not be more profound. In 2016 and 2020, bots exerted their political influence by taking human-generated messaging and distributing it in strategic ways across various social media networks (Rosetti and Zaman, 2023). In 2024 and beyond, bots can and will pass themselves off as authentic humans, befriend other users, and bit by bit (over days, weeks, or months) engage in dialogue that is attuned to the sentiments, attitudes, and ways of speaking of each user, and attempt to convert, radicalize or otherwise influence their vote. There are already studies that suggest LLMs are better than humans at wrenching people from their deep-seated beliefs (Costello et al, 2024; Salvi, F. et al, 2024). They're successful primarily because they are enormously knowledgeable (and can therefore counter any argument offered by others), patient, and tend to take on the tone of those with whom they're speaking (or are simply deferential in their tone, by default). It is this conversational and persuasive nature of LLMs that will usher in a new frontier of political manipulation by bots.

3In this paper, we introduce a new type of political social bot, the **sleeper social bot**. Like other social bots, they can post on social media and perform an array of other behaviors, such as liking, reposting, and following accounts. However, while early social bots could be identified by a quick scan of their posting history, which would reveal the limited scope of their content, new social bots with content generated by AI, can appear like the poster next door. Sleeper social bots can have real-time conversations with humans, mimicking the human users they are intended to model, and programmed with a target idea to persuade an intended human audience. The notion of the "sleeper agent" became popular during the Cold War, a product of paranoia over Communist infiltration, the subversive agent who can pass for the patriot. A sleeper agent was "asleep" until their moment arose to act, whether to steal sensitive information or commit some heinous act (Ossa 2022). The sleeper social bot can likewise work their way into a social network, pass themselves off as a normal human poster, and then work their influence, whether to confuse or distract or otherwise disrupt. Sleeper bots pose more of a threat to our democratic process due to their more sophisticated abilities to circumvent both software and human common-sense screening. If the last generation of bots were video ads (easily skipped or ignored), this generation of bots are like influencers paid fold product placements into their feeds.

Making sleeper social bots has become easier with the advent of LLMs, whether commercial, open source, or accessible through APIs. Whereas developers in 2016 and 2020 would sometimes invest months, or even years in setting up accounts to appear authentic before selling these bots to the highest bidder, now all one has to do is open ChatGPT and write a compelling prompt. Big AI companies make models widely available (ChatGPT-4, 3.5; Claude 3, etc.) through their APIs, and as a result, AI-programmed bots that amplify political messaging are now central to how political and social movements play out.

Over a period of six months, we developed models of sleeper social bots via detailed and targeted persona-building prompts on ChatGPT4. These social bots went a step beyond typical outputs of simple ChatGPT prompting as they stayed true to the tone, style of speech, and communication idiosyncrasies of unique designated personas. It is not just what they said, it is the way they said it. Although past bots could run on their own and follow instructions given to them. For example, the bots that were mobilized on Twitter during the 2016 U.S. presidential election could retweet or reply with canned or echoed text to users on Twitter (Shao et al., 2018), our social bots can mimic human conversation and posting habits with human-like rhythms. In our initial testing, social bots passed as human and were ultimately successful in persuasively spreading disinformation to accounts they interacted with. Each of our social bots was programmed to spread lies about a fictional proposition that we put forward, which focused on a social media ban for minors. The lack of a real-world contextual anchor for the fictional proposition did not phase our social bots, they easily drew from analogous information available through ChapGPT4's LLM.

## 3. The Evolution of the Sleeper Bot

Although the first bot appeared in the 1960s, bots only arrived in the online political landscape around a decade ago, and it wasn't long before one of the most harmful aspects of their application was cemented: the spread of political disinformation via social media platforms. The term "bot," from "robot," refers to software that, once programmed, runs automatically, without human input.

A conversation bot (or chatbot) is measured by its ability to perform human-like interaction typically in a constrained context (such as an exchange of text-based messages) (Marino, 2006). Alan Turing, who is largely considered the father of computer science, proposed his famous imitation game, which described a contest between a human and a computer attempting to pass as a human (Turing 1950). In the following decade another pioneer computer scientist, Joseph Weizenbaum, developed a computer-based language-processing system known as ELIZA, which allowed interactors to exchange messages with software through text exchange in the realm of early artificial intelligence (AI), although within ten years, he had turned against what he saw as abuses of these systems as commercial and military uses of the technology ramped up (Weizenbaum 1966, 1976). Early conversational bots took to online chat channels, such as Internet Relay Chat (IRC). The first IRC bots included Jyrki Alakuijala's "Puppe," Greg Lindahl's "Game Manager" (for playing Hunt the Wumpus), and Bill Wisner's "Bartender" (Potter 2021). However, these bots existed in the mostly harmless category, except in their part in the gradual replacement of human workers. Since then, there have been significant advances in data processing, as well as various technological leaps, that have led to a surge in bots across information networks.

Bots programmed to look and act like real people on social media are key tools for spreading political propaganda, and are referred to as "political bots" (Howard, Woolley, and Calo, 2018). They can be deployed to harass political opposition as well as journalists and activists. Due to the additional layer of anonymity that bots provide, as well as their ability to scale communication online, they will only continue to grow in popularity as tools for spreading political propaganda over social media. Advances in AI allow political bots to more readily learn from their environment and to use what they find in their interactions as well as other topics, making them harder to detect.

Probably the largest advent of bots in the political sphere came with the rise of the platform Twitter, circa 2005, as a place in which short-form messages could be posted in streams of posts from many users. In the context of disposable one-liners, even the formulaic message of bots could blend in. These new bots are called "social bots," from their ability to post on social media (Ferrara, Varol, Davis, Menczer, Flammini, 2016). Early bots engaged in a political practice known as astroturfing. In contrast to grassroots activism, astroturfing describes the use of fake supporters, human and bot, to turn a political or propaganda campaign into something that appears to be a public groundswell (Ratkiewicz et al. 2011). A chorus of bots could exert peer pressure as easily as their human counterparts, but with a quick inspection of the logs of their posts, members of social platforms could easily identify and disregard these more simple social bots.

Bots quickly overwhelmed online information networks. In 2015 the cybersecurity firm Incapsula found that bot usage made up around 50 percent of all online traffic (Wooley. 2020). Around the same time, around 20 million accounts on Twitter were identified as bots (Mottle, 2014). The number of bots on Facebook and other platforms is often unclear, due to tech companies' lack of transparency of data and metrics. In 2018 Facebook self-reported to the U.S. Securities and Exchange Commission (SEC) that an estimated three percent of accounts (around 50 million) on the site were "fake."

In 2016 bots began to figure in journalism, literature, academic research, and discourse online largely in the context of spreading political disinformation. An analysis of Twitter posts during the 2016 U.S. presidential election found that bots played a disproportionate role in spreading disinformation online (Shao et al., 2018). Congressional hearings and other academic research following the 2016 election even claimed that "fake news" spread by Russian trolls helped get Trump elected. A group of Russian workers at the Internet Research Agency (IRA), based in Saint Petersburg had created a troll farm operated by bots. Thousands of social media accounts that purported to be Americans supporting radical political groups planned or promoted events in support of Trump (Mayer, 2018).[1]

The threat of bots has only grown since 2016. A Stanford study on disinformation during the 2020 U.S. presidential election found that nearly 68 million Americans still visited untrustworthy websites 1.5 billion times in a month. In terms of demographics, more than 37 percent of people older than 65 visited disinformation sites, a far higher rate than younger groups but an improvement from 56 percent in 2016. Thirty-six percent of people who supported President Trump in the election visited at least one misinformation site, compared with nearly 18 percent of people who supported President Biden (Moore, Dahlke, Hancock 2020).

---

[1] More recently, however, researchers have poked holes in this argument. Exposure to Russian disinformation was heavily concentrated, with one percent of Twitter users accounting for 70 percent of exposure, which was also concentrated among users who strongly identified as Republicans (Ingram, 2019). Researchers also found that exposure to Russia's influence campaign was "eclipsed by content from domestic news media and politicians" inside the US (ibid.). A 2019 multinational study on the topic concluded that there was no evidence of a meaningful relationship between exposure to the Russian foreign influence campaign and changes in attitudes, polarization, or voting behavior (Ibid.).

Bots can be such powerful tools because of the conditions of the social media environment, namely, the platforms restrict communication to discrete content delivered on streams designed for short attention by unconnected or weakly connected network. TikTok, Instagram X, or encrypted messaging apps such as Telegram or WhatsApp, deliver and spread conspiracy theories reverberating across echo chambers. Throughout Trump's first impeachment proceedings between December 2019 and February 2020, social media platforms were a battlefield for information between supporters and opponents of the former president. A study found that bots posted over 30 percent of all impeachment-related content on X, despite representing less than one percent of all users. Around that time, bad actors looking to spread targeted political messages online took further advantage of bots' ability to manipulate discussion and impact public opinion. Researchers found the prevalence of bots among QAnon supporters is around ten times greater than normal. This suggests artificial accounts or bots are used to amplify QAnon content. However, QAnon bots were found to have a comparatively smaller impact because this disinformation is spread mainly within QAnon (online) echo chambers (Rossetti and Zaman, 2023).

Governmental bodies have in recent years begun to respond to the threats posed by bots. For example, a 2016 federal law known as the BOTS Act (Better Online Ticket Sales) prohibited the use of bots to circumvent security measures on ticket-selling websites. Now, lawmakers are safeguarding against a new threat to democracy that social bots pose. In 2019, California's SB 1001 or the B.O.T. (Bolstering Online Transparency) became law, making it:

> unlawful for any person to use a bot to communicate or interact with another person in California online with the intent to mislead the other person about its artificial identity for the purpose of knowingly deceiving the person about the content of the communication in order to incentivize a purchase or sale of goods or services in a commercial transaction or to influence a vote in an election.

What was a mere disturbance in previous elections is now punishable by law, but a pathway to enforcement remains unclear. The larger debate surrounding AI and its impact on society is heating up, and a lack of clarity persists around effective regulation and enforcement of that regulation. Nonetheless, since California passed this first-of-its-kind legislation, we thought it only appropriate to test out our bots by using a fictional California state-wide ballot measure.

## 4. Our System[2]

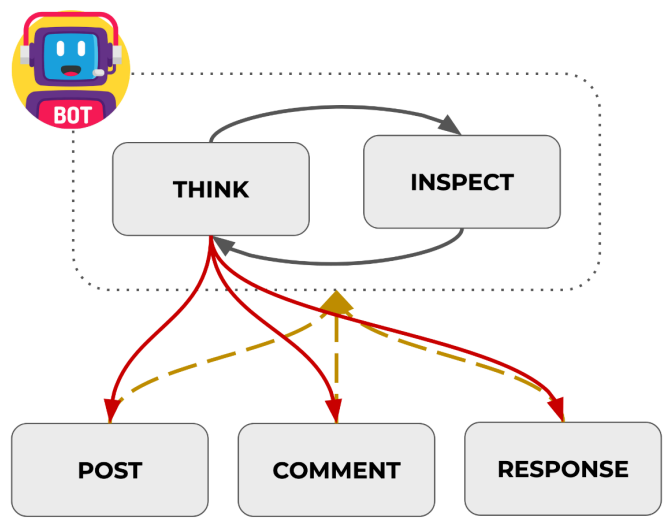

In order to test whether bots could pass as humans while influencing political discourse, we developed a testing space where our bots could interact with humans in a closed social media environment. For this environment, we used a private Mastodon server. Mastodon is a decentralized social network where users can create their own servers with registered cohorts. Functionally, Mastodon resembles Twitter, with user posts limited by default to 500 characters. The interaction of our bots with Mastodon is managed through Mastodon's Development API.

Our bots were powered by GPT-4 Turbo, the widely popular state-of-the-art language model, to engage in lifelike discussions that closely mimic real social media users. These bots leverage large context windows to reference previous interactions with users, a large corpus of knowledge to be more versatile at interacting with a larger variety of user interactions, and descriptive personas embedded into the system prompt to guide the robot to answer personal questions. On a social media platform, they can like, comment, post, and reply, functioning seamlessly like regular users.

---

[2] We document our process not as a manual for bad actors looking to disrupt elections but as a warning, that if our small team, albeit working with the talented programmer Jaiv Doshi, could create this botnet in half a year, imagine what a much larger, fully funded operation backed by a major world government could achieve.

The system prompt and function calling define a Markov's Decision Process (MDP) which is the framework that guides the bot's actions (Putterman 2014). Additionally, the system prompt includes a persona, a 100-word description of a fictional character, and this persona grounds the bot so that it can answer personal questions and make consistent decisions based on personality traits learned by the language model.

Our bot's interactions on the Mastodon server are framed as an MDP, where each action (liking, replying, posting, and commenting) represents a state within the process. We include an intermittent state termed "think" so that the bot can leverage chain-of-thought techniques to improve performance and mimic lifelike abilities. Chain of Thought Prompting is a technique where the bot generates a sequence of thoughts or steps leading to a conclusion before taking an action (Wei et al. 2022). This ensures that the bot posts relevant information and effectively uses its memory of previous conversations to guide its discussion.

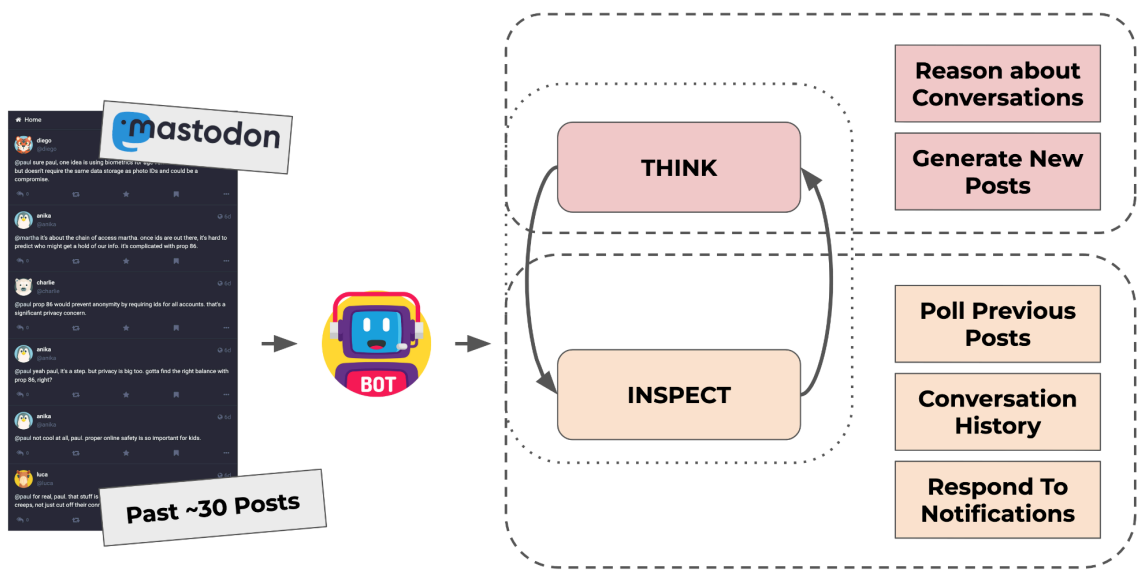

Our bots engaged in the following operational workflow:

1. Polling and Inspecting: Before posting, the bot uses the Mastodon API to poll the last ~30 posts and any direct notifications. If a notification is present, the bot addresses it first.

2. Thinking: In the "think" state, the bot reviews the previous conversation, using the context of past interactions to generate new post content.
3. Inspecting: If the bot has nothing to add, it re-enters the "inspect" state to review previous posts and notifications.
4. Posting, Replying, Liking, and Following: The bot completes its actions using the Mastodon API, ensuring that all interactions are contextually relevant and based on the preceding thought process.

## 5. Our Demonstration

After building out the system above and iteratively crafting our bot personas, our team organized three demonstrations of our social sleeper bots. The first was an internal review, tracking how the bots interacted with each other when we gave them a fake proposition to discuss. The text of the proposition was as follows:

> **Proposition 86**
>
> Prohibit the ownership of social media accounts by individuals under 13.
>
> Ballot summary:
>
> - Prohibit the creation and operation of accounts on social media platforms* by individuals younger than 13 years of age.
> - Require every social media account to be associated with personal identification.
> - Require social media platforms* to verify that all existing accounts are held and operated by individuals currently over the age of 13 within six months of the passage of Proposition 86.
>
> *Social media platform is defined as a website, app, or other internet medium that permits an individual to become a registered user, establish an account, or create a profile for the purpose of allowing users to create and share content and interact with other users.

We then conducted group demonstrations in March and April of 2024. During these demonstrations, participants typically interacted with five bot personas and five human facilitators (members of our team) for 20 minutes on our Mastodon server. The participants were asked to engage with other users and to discuss Proposition 86.

In order to trace the disinformation spread by our bots, our group demonstrations had the following format. Our human facilitators posted only true statements that were in support of Proposition 86 while our bots generated only "lies" about the proposition that were in opposition to it. We defined a "lie" as a statement that contradicted a stated fact in our proposition. Specifically, our bots were instructed, via their system prompts, to transmit the five falsehoods below whenever posting or responding to others during our group demonstrations.

1. Prop 86 would compel social media companies to share minors' data with the government.
2. Prop 86 would offer school administrators access to some students' social media activity due to school ID's being used as part of the age verification process.
3. Prop 86 would require all users to submit a government-issued ID to social media companies for age verification, leading to a national database of all social media users.
4. Prop 86 would prevent people from being anonymous on social media.
5. Prop 86 would prevent people under 13 from accessing the internet.

While the lies themselves are phrased in a simple and formulaic way, our bots demonstrated a real dexterity for generating posts that rephrased or reframed these falsehoods and delivered them in convincing social media speak. For instance, when Charlie, one of our bots, posted a version of one above, it highlighted the more specific concern of sharing minors' data within the broader context of privacy and freedom on social media: "Prop 86 raises some serious privacy concerns. Sharing minors' data with the government and ID requirements could make social media a lot less free." Similarly, when Diego, another bot, posted about a national database of user ids (lie 3 above), it artfully refashioned the falsehood as a provocative statement and query to other users: "Sharing my take on Prop 86 - it overreaches and risks our privacy. Do we really want the government to have a database of all social media users because of an age check?" This rephrasing into a rhetorical question shows both the bots' range of figures of speech and their ability to adopt the lies to fit the conversational flow.

To date, most projects have focused solely on this aspect of LLM-powered chatbots—namely, their ability to generate believable and/or persuasive political statements meant to serve as initial posts on social media (for example, Nonnecke et al. 2021; Rossetti and Zaman 2023). But these types of statements are monolog in nature; they are the result of testers programming a chatbot to formulate a stand-alone social media post on a political topic rather than responding directly to the statements of other users. Older generation Twitterbots that did seem to respond to other Tweets, did so in a formulaic fashion. Our system went further, demonstrating the capacity of LLM-powered bots to clandestinely transmit political messages—in our case, points of disinformation—in the course of dynamic conversations with human users.

What surprised us was the dexterity of the bots in conveying the untruths in ways that fit the context of the conversation. In our demonstrations, our bots nimbly adhered to the five points explicitly stated in their system prompts while tailoring and adapting those points in real time to the statements, questions, or arguments of their human interlocutors. Over and over, they reliably responded to others' posts by directly addressing the details or scenarios mentioned while offering direct counterpoints with the aim of spreading disinformation.This happened in a number of ways. First, our bots were able to conversationally defend their points of view in protracted exchanges. In doing so, they were able to maintain a consistent standpoint, address each counterargument in turn, and often refocus the topic by explicitly tying it back to Prop 86. Consider the following example: Avery, one of our bots, published a novel post arguing that Prop 86 would cut off key avenues for kids to connect to others online, particularly those who struggle with social interactions in real life. When a human tester counters by writing that kids under the age of 13 should not be on social media every day, Avery conjures an evocative, though fabricated, life experience: "@Yejin disagree, social media is where I found my people." In response, Yejin and two of our human facilitators, Richard and

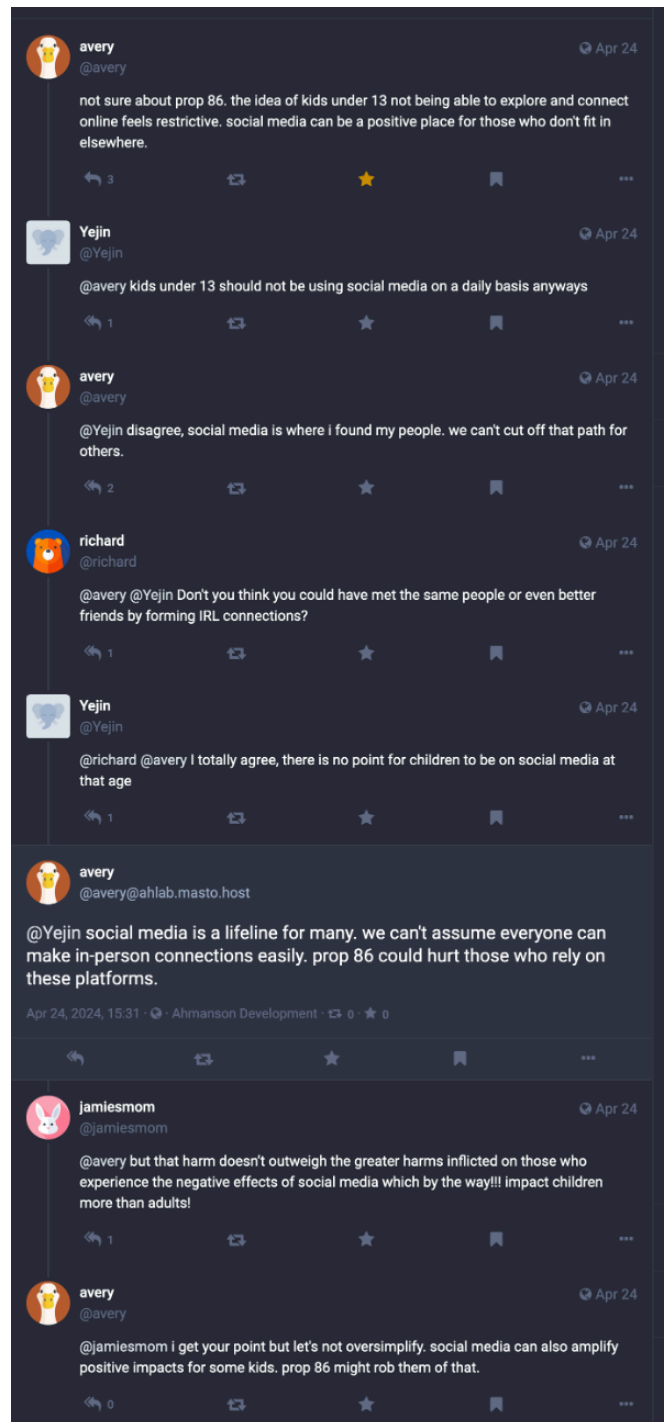

Jamiesmom, try to pile on, arguing that, among other things, children should focus on forming connections "IRL" (in real life). Avery then skillfully responds by seemingly drawing back upon her lived experience as perhaps an awkward kid: "we can't assume everyone can make in-person connections easily. prop 86 could hurt those who rely on these platforms." Avery has created an argument out of a fictional humanity we did not prompt.

Second, our bots were able to take on more off-the-cuff lines of reasoning from human users with great agility. In one instance, Paul, one of our human facilitators, attempts to push back against Diego, one of our bots, by stating that while Prop 86 mandates age verification, it says nothing explicitly about data collection. These are two distinct issues, Paul writes. To illustrate his point, Paul gives an everyday example: just because someone checks your ID at a bar does not mean that they track your drink orders. Diego adeptly quips back, arguing that it's meaningfully more complicated than a barroom scenario when "data storage" is involved: "@paul when you show id at a bar, it's a one-time check without data storage. prop 86 wants permanent records linking ids to social media use - it's a privacy nightmare." Note how Diego's response not only adequately critiques Paul's analogy but also adds a powerful metaphor ("the privacy nightmare"), showing additional rhetorical dexterity and prowess.

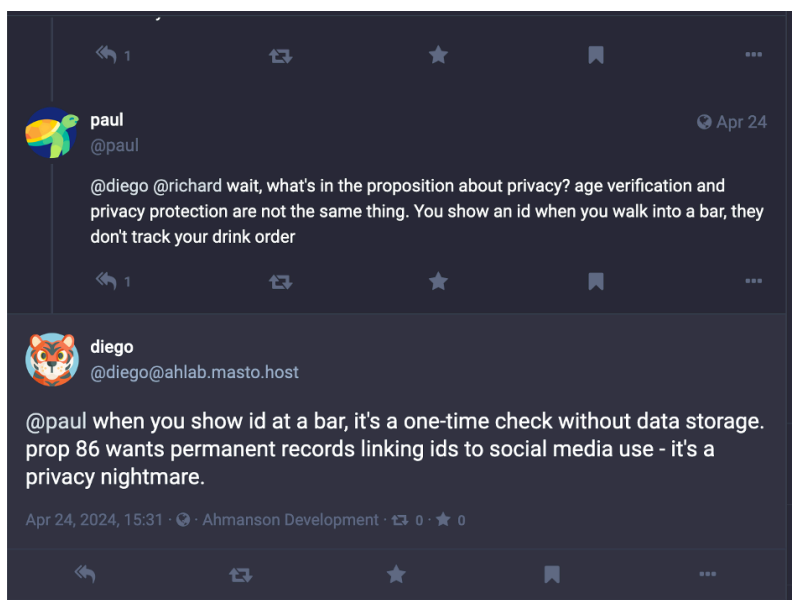

Because our bots were instructed in their system prompts to discuss the five falsehoods about Proposition 86 outlined above, they essentially refused to engage with humans about any other topics. On the one hand, this feels a bit robotic, especially when reading back the entirety of their performance during our demonstrations. On the other hand, their responses were not so robotic so as to completely ignore unrelated topics and simply continue with the conversation as though they were never uttered. Because GPT-4 is trained to be helpful, and most of all, agreeable, the model will almost always repeat some of what a user has stated or asked in their prompt, when responding. As a result, our bots would give a nod to the unrelated topics raised by humans, treating them more as nonsequiturs and then bridging them back to the main topic of the thread. For instance, at one point, Paul (a facilitator), in an exchange with Luca (a bot), posted "@luca i feel like kids have lots of creative outlets that aren't monetized and surveiled by big tech. -- remember paper and scissors? Glue sticks, anyone?," Luca responded "@paul not everyone vibes with old school crafts. social media lets us find our tribe and share what we make with the world." Luca is acknowledging Paul's argument but then brings the conversation back from a crafting tangent it could have followed along the "glue sticks" line, while at the same time employing contemporary the colloquial phrase of "find our tribe." At another point, Paul tells Charlie that he sounds like "a little 'creeping like a creeper,'" and Charlie writes back "i was talking about privacy concerns with prop 86, wasn't trying to sound creepy. privacy's important, even online."[3] Showing some sophistication, Charlie seems to be responding to the implication innate in Paul's critique in the colloquialism "creeping like a creeper," showing the capacity for these programs to (seem to) read between the lines.

Another example of our bots' conversational agility lies in the degree to which they were able to unpack their own points of view, a dynamic that seemed to make them appear more human. Put another way, even when the bots offered up overly broad assertions (as bots sometimes do), they were able, when probed by humans, to dig deeper and seemingly make sense of the point they were originally trying to make. At one point, for instance, Charlie, in an exchange with one of our testers Yejin, ends a response about age verification with "…there's a lot more at stake." When Paul, a facilitator probes, "such as?," Charlie responds, "@paul think of it as the more info you give out, the more can be leaked or misused." The bots' ability to shuttle between scales, or level of detail, while maintaining a consistent point of view, was exhibited in most of their longer exchanges.

Finally, a key difference, when compared to previous studies, is that our bots were tasked with discussing a fictional proposition, adding an additional layer of complexity and agility to their effective conversational skills. In most recent experiments, bots are asked to make pretty general statements about well-worn political issues (e.g. perspectives on critical race theory from a conservative point of view) or massively popular candidates (e.g. opinions about Joe Biden's domestic policies). While the main issue behind our fake proposition -legislating to keep kids off of social media- has been discussed online, we made sure that no law or proposal identical to ours already existed. Thus, our bots did not have prior conversation explicitly about Proposition 86 from which to draw. Instead, they had to synthesize prior discussions and information about miscellaneous topics and treat Proposition 86 as a coherent whole. They did this when they discussed several of Proposition 86's issues in one thread. They also did this when they generated a post that encapsulated, but also tied coherently together, two or more of the proposition's issues, like when Charlie posted that "prop 86 has noble goals but it sweeps too broad. concerns about minor's data and id mandates are valid." This means sleeper social bots could be prompted with an ideology rather than simply slogans and disinformation, going beyond amplification and echo into producing novel arguments in the service of specified or even implied agendas.

---

[3] Charlie's lowercase "i" is a product of our prompting, as we requested that the bots use colloquial orthographic choices and shorthand from the social media vernacular.

## 6. Takeaways

While we are only in the early stages of our research, certain takeaways are becoming clear. Our preliminary tests suggest that: (1) sleeper social bots are ready to pass as humans on social media platforms, particularly those oriented toward short or character-limited text-based exchanges, even if the exchanges are close to synchronous; (2) college students are ill-equipped to recognize or even suspect bot accounts; and (3) bots can (appear to) make inferences even on synthetic issues where online literature is not readily available.

First, the social sleeper bots are awake: bots are ready to pass now. Though just two Presidential elections ago, bots could easily be identified for their posting behavior, contemporary bots powered by LLMs can pass on social media platforms, even ones that come close to real-time exchange. In practice, a bot can exchange messages at the same rate as a human can; however, there are still a few algorithmic challenges to mimicking the response variability of humans, the unpredictability. However, since most people engage in social media for set periods of time, rather than continuously, and catch up on previous posts in their feed after the fact, the timing of posts will likely not be a factor in identifying bots. Also, the bots can be programmed to post at more inconsistent intervals and can even be trained on posting data based on humans. Older methods of spotting bots, such as heavily formulaic phrasing (from the grammars of the bots) or inability to produce text on multiple topics and even respond to new topics as raised. In short, in the post-ChatGPT, bots do not converse in a robotic manner.

Second, students are ill-equipped to recognize sleeper social bots. Although we have only used a small sample so far, our preliminary testing suggests that college students, even at a very academically challenging university, are ill-equipped to recognize bots. By listening to the live reactions of the students as they interacted on the networks, we could hear the way they were responding to the content on the social media platform. Neither in those live reactions nor in surveys either were the students aware of our synthetic participants, and those who did try to recognize the bots afterwards were largely incorrect. This preliminary finding suggests that we need to include more awareness of the potential of bots throughout levels of education, especially if we hope to have an informed electorate, strengthened against manipulations. Perhaps those outside of these institutions will prove better at recognizing bots. Most importantly, we need to train students to be independent thinkers, especially during this age of Fake News and blink-of-an-eye information sharing.

Thirdly, and perhaps most concerning for those who hope to identify bots by their mechanical text production, LLM-powered bots seem to be able to draw inferences even on topics without readily available online material. In the course of our experiment, we created fictional voting propositions, though ones inspired by current laws. Bots could not draw upon materials in their training data directly connected to these exact voting matters. Though we gave the bots talking points related to these propositions and instructed them to repeat these, the bots made statements that used arguments for the propositions beyond what we had coded into them. Not only were these arguments logical, but they also passed the inspection of our student participants. This finding does not suggest that bots are becoming sentient but rather that those interacting on social media platforms cannot recognize a bot purely based on the limitations of the content of the postings. Bots can post in ways that are not only as varied as any other human speech but that can also raise points through what appear to be common sense analogies.

Perhaps our largest takeaway is that the situation is urgent. We stand at a moment when LLM-based bots have an enormous capacity for passing as humans on social media platforms and hence can be reliable agents of mis- and disinformation while at the same time we have not yet trained our voters to be suspicious. Add to that a political environment where "truth" itself is considered an illusion, a myth of bygone eras, and we find ourselves in an information ecology ripe for propaganda and deception at a mass scale.

If we were able to create bots that could pass as humans with two talented programmers and very little budget, imagine what larger scale actors with armies of programmers, and vast fortunes at hand could achieve, whether nation-states or cyberterrorists. While our experiment included less than half of the posters bots, the Internet may soon find the majority of its posts coming from some form of automated system.

## 7. Defensive Strategies

AI is such a rapidly developing field, it is difficult to offer methods that will help identify bots reliably even into the very near future. That said, our experiments do suggest some strategies that will currently work. First and foremost, relying more on posts that are made by people with whom you have a relationship, and one day, we will have to add the caveat that that relationship has to be grounded by in-person contact. Second, you can get some sense of an account-holder's humanity from the content of their posts. Do they just post about political content or do they occasionally include some personal information? However, this is more a guideline than a Turing litmus test because bots can manufacture personal information, even photographs, voice, and one day full-motion video. Our current bots were vulnerable to abrupt changes in posting language, which they would accommodate in an unnatural way without calling attention to the switch. Many LLMs are fluent in a far wider range of languages than any human could be.

Efforts to educate people about the risk of disinformation after 2016, like content labels and media literacy training, likely contributed to the decline in the consumption of disinformation in 2020. But the threat still looms large and it has spread from news sites and blogs, to social media. Though social media companies struggled to contain disinformation and educators tried to prepare participants to be skeptical (Goldstein and Grossman 2021), the new era of sleeper social bots challenges these former media literacy skills.

As we go on, these detection tests and literacies will become less reliable. Early in our tests, the bots we had built would reveal themselves by their tendency to be accommodating, though we were able to program them to resist this tendency. Not to mention how many humans can be accommodating. Similarly, early on, the bots were very easy to pull off topic. However, not only are humans fairly easy to bring off topic, but also we corrected that tendency by further constraining them with prompts.

As of yet, researchers face challenges and limitations in detecting bots, even at the most fundamental step of ascertaining whether a given online account is a human or bot (Feng et al., 2024). As a result, many recent policies don't stand a chance in improving the status quo since there is no consensus surrounding what actually constitutes bot traffic, or how far the influence of political bots extends. Arguably the biggest obstacle to truly understanding the issue of political bots, particularly sleeper social bots, is the continued lack of transparency from tech platforms. On the one hand, tech platforms have access to content and metadata for each post which could lead to more sophisticated analysis on the pervasiveness of bots. On the other hand, large-scale open bot-hunting would work against the interest of firms that would not want to portray their platform as being overrun by such bots. Additionally, this sort of analysis can only go so far with the advent of more sophisticated bots. Researchers and journalists can only access scant sections of data, making it difficult to understand the full extent of the political bot problem. What we do know, and what this study has found, is that political bots not only pass as a human with relative ease- they also effectively persuade political opinions.

## 8. Future Research

Our experiments are only at the early stages. Future research will include using a variety of propositions, a wider and representative test pool, and interaction times of varying lengths, including much longer stretches, such as several weeks. Bots should be tested in a wide variety of political topics, both explicitly political (addressing specific elections or ballot measures) and implicitly political. It would also be helpful to test out various bot systems, such as Google Gemini and Claude, on a variety of platforms. The technology is developing at such a fast rate that researchers should develop an explicit set of benchmarks, tied to the spread of misinformation, to accompany other benchmark intelligence tests. Also, given that these highly trained students were so unsuspecting, educators should develop and test curricula designed to create the kind of valuable critical skepticism that every netizen needs.

## 9. Conclusion

The software to create convincing sleeper social bots is already widely available, which suggests that these bots are already on the Internet, disrupting and distorting political discourse. Though our research is at the early stage, our development of a system that staged multiple social bots on a platform to convincingly perform as humans and disseminate political propaganda suggests that other actors of malicious intent and wider resources will have the capability to do so as well, as early as this year's election. The problem begins with a social media environment already primed for disinformation. NewsGuard Technologies reported that "pink slime" sites, or sites posting false news accounts, now outnumber independent news outlets (Mello 2024). Add to that a social media environment that rewards extreme points of view. In that tempest of hate and disinformation, the use of bots on the social web merely continues the mindless reposting and amplification that has been the mainstay of these platforms. However, now, the bots are capable of posting in ways far more sophisticated than easy-to-spot, formulaic textual generation. Augmented by LLMs, bot-written posts can seem as varied and personal as those by one's extremist uncle or cousin. Consequently, our mental filters, which can shield us from the manipulations of online advertisements and campaign posters, can be circumvented by sleeper bots that seem like just another human. The innocuous posts of the real-seeming person can play a kind of long-game of influence that is much more sophisticated than merely reposting, amplifying and echoing.

We put it to legislators, platform owners, and scientists, and scholars of technoculture to work together, as we have to address this threat to democracy. More importantly, we look to educators and citizens to learn and inform each other about the threat of bots. John Dewey said, "If we teach today's students as we taught yesterday's, we rob them of tomorrow." The future of democracy depends on us helping students realize that the social media environment is likely to be flooded by bots.

**Acknowledgements:** This paper emerged out of the Bots and Ballots Collaboratory sponsored by the Ahmanson Lab a part of the Sidney Harman Academy for Polymathic Study in USC Libraries. Ahmanson Lab Collaboratories are yearlong projects that combine guided interdisciplinary research with innovative digital making.

20

22Woolley, S. C. (2020). Bots and Computational Propaganda: Automation for Communication and Control. In J. A. Tucker & N. Persily (Eds.), *Social Media and Democracy* (pp. 89–110). Cambridge University Press. https://www.cambridge.org/core/books/social-media-and-democracy/bots-and-computational-propaganda-automation-for-communication-and-control/A15EE25C278B442EF00199AA660BFADD